# Ensemble learning and iterative training (ELIT) machine learning: applications towards uncertainty quantification and automated experiment in atom-resolved microscopy

Ayana Ghosh,[1,2] Bobby Sumpter,[1] Ondrej Dyck,[1]

Sergei V. Kalinin,[1,a] and Maxim Ziatdinov[1,2,b]

[1] Center for Nanophase Materials Sciences, Oak Ridge National Laboratory, Oak Ridge, TN 37831

[2] Computational Sciences and Engineering Division, Oak Ridge National Laboratory, Oak Ridge, TN 37831

Deep learning has emerged as a technique of choice for rapid feature extraction across imaging disciplines, allowing rapid conversion of the data streams to spatial or spatiotemporal arrays of features of interest. However, applications of deep learning in experimental domains are often limited by the out-of-distribution drift between the experiments, where the network trained for one set of imaging conditions becomes sub-optimal for different ones. This limitation is particularly stringent in the quest to have an automated experiment setting, where retraining or transfer learning becomes impractical due to the need for human intervention and associated latencies. Here we explore the reproducibility of the deep learning for feature extraction in atom-resolved electron microscopy and introduce workflows based on ensemble learning and iterative training to greatly improve feature detection. This approach allows incorporating uncertainty quantification into the deep learning analysis and also enables rapid automated experimental workflows where retraining of the network to compensate for out-of-distribution drift due to subtle change in imaging conditions is substituted for human operator or programmatic selection of networks from the ensemble. This methodology can be further applied to machine learning workflows in other imaging areas including optical and chemical imaging.

[a] sergei2@ornl.gov
[b] ziatdinovma@ornl.gov

**Introduction**

Electron and scanning probe microscopies have emerged as primary techniques for exploring the micro-, nano-, and atomic scale worlds.[1-3] Multiple examples of the imaging of materials classes ranging from metals and semiconductors to biological and macromolecular systems are abound, with the microscopic tools becoming the linchpins of academic and industrial laboratories throughout the world.[1, 4-6] This rapid progress in imaging techniques have further transformed many imaging areas from largely qualitative to quantitative. Traditionally, in atomically resolved scanning transmission electron microscopy (STEM) or scanning tunneling microscopy (STM) the attention of the researcher has been focused on the presence of large-scale morphological features such as surfaces and interfaces, localized or extended defects, with the conclusions on the physics and chemistry of materials driven by these qualitative observations. Comparatively, the progress in high-resolution imaging allowed quantitative information on the materials structure to be obtained, including the positions of the atomic nuclei in STEM, center of mass of electronic density of states in STM, etc. This information in turn is related to the fundamental physics and chemistry of materials, and several examples of quantitative studies of materials physics from atomically-resolved quantitative observations are now available, including mapping of polarization fields,[7-12] octahedra tilts,[13-16] and strains in (S)TEM[17, 18] and surface distortions in STM[19].

However, this progress necessitates rapid analysis of the individual images, as well as dynamic data sets obtained during, e.g., temperature,[20, 21] environment,[22-24] or beam-induced transformations[25-29] of solids. The goal of such analysis is to transform the data stream from the microscope, *ex-situ* or *in-situ*, into the coordinates and trajectories of specific features.[30] These features can be atoms and molecules in the atomically resolved techniques, or larger scale objects such as nanoparticles and nanorods, cells, etc., in mesoscale imaging. Once established, these features can be used for exploring relevant physics, e.g., strain mapping, visualization of order parameter fields such as polarization or octahedra tilts, and subsequently deriving generative physical models. The need for such analysis necessitates two additional aspects. The first is uncertainty quantification, i.e., ascribing "a degree of trust" to the experimental results. The second is the latency of the analysis, a consideration that becomes particularly important in conjunction with autonomous experimentation[31-33] and electron beam atomic fabrication.[27, 28, 34-37]

Traditionally, this analysis has been implemented using a broad variety of image analysis tools, ranging from simple maxima finding to correlative filters to Hough transform based techniques, with each domain area evolving a specific set of tools. Most of these methods required extensive tuning of the hyperparameters in the beginning of analysis cycle, and often require operator input throughout the process. As such, analysis of more than several images has been unusual. The breakthrough in this area, as in many other areas of computer vision, is the broad implementation and availability of deep learning (DL) networks, pioneered by AlexNet[38] and evolving to the large families of ResNets,[39] U-Nets,[40] etc. In STM, the application of DL for image analysis was pioneered by Ziatdinov *et al.* for molecular resolved imaging[41] and Wolkow for atomically resolved imaging,[42] and in STEM by Ziatdinov et al.[43] This initial effort has grown exponentially in the recent years.[44-46]

However, applications of DL for atomically resolved imaging, or, equivalently, discovery of large number of similar objects, has significant differences from the classical DL applications. Traditionally, DL methods are optimized to perform feature finding on a large number of possible classes, such as 10 digits in MNIST database or 10 image categories in CIFAR data bases, with multiple examples in each class. Similarly, these are associated with large variability within the class. Comparatively, atomic and particle finding problems typically necessitate finding (almost) identical objects, whereas the imaging conditions can vary between dissimilar experiments or between simulations and experiment. Consequently, the applications of DL in experimental data analysis have to deal with significant out-of-distributions effects. Retraining of the network for a different parameter set is time consuming, both on the labeling and training side. This aspect is particularly limiting for an automated experiment where analysis must rapidly adapt for changes in imaging conditions. At the same time, a model trained to account for a broad distribution of experimental parameters may fail to recognize multiple subtle atomic features within a single experimental dataset. Additionally, the scientific applications of DL require meaningful uncertainty estimates, which has been a challenge for applications of classical DL to real-world data.[47]

Here, we introduce the iterative DL workflow that surpasses these limitations. This approach utilizes the heavily degenerate nature of high-resolution microscopy, in which relatively small number of feature classes are possible, and even within a single image multiple realization of the class are present. We also utilize the fact that a deep learning a model's final state can be very sensitive to random initialization of weights and shuffling of training mini-batches. These two effects are harnessed in a workflow combining ensemble learning (EL) by multiple neural networks, allowing for selection of artifact-free features and pixel-wise uncertainty maps, and iterative training (IT) where the discovered features are used to retrain the network, focusing its attention on features present in the (heavily degenerate) data and thus increasing the detection limit of the network on the dataset(s) of interest. This allows revealing previously unrecognized features, and to compensate for out-of-distribution drift during the experiments.

**Results and Discussion**

The classical DL workflow consists of preparing a single labeled training set, selecting appropriate neural network architecture, splitting the prepared training set into 2 (train + test) or 3 (train + test + "holdout") parts, and tuning the training parameters until the optimal performance on the test and/or "holdout" set is achieved. Once trained, the neural network is expected to generalize to new, previously unseen data. In the experimental sciences, where the labeling process typically requires an in-depth domain expertise and as a result the availability of labeled data is usually very limited, one can sometimes use ab initio simulations to prepare a training set. The common challenge for real-world applications of DL (that is, application to a stream of new unfiltered data beyond the test and "holdout" sets) are the out-of-distribution effects such as changes in data acquisition and processing (inside of an instrument) parameters. An example is the failure of state-of-the-art DL models trained to detect pneumonia in chest X-ray data to provide accurate results on new hospital's data due to the small variations of instrumental and data acquisition parameters.[48]

To set the context we consider the application of DL to experimental imaging of atoms. For DL-based image analysis, the labeling of training data can be performed on the level of individual images (image0001 is "structure A", image0002 is "structure B", ...., image6999 is "structure B") or individual pixels (a pixel at coordinate ($i_1$, $j_1$) belongs to "structure A", a pixel at coordinate ($i_2$, $j_2$) belongs to "structure B", etc.). The latter is referred to as a DL-based semantic segmentation and will be the focus of our work. Here we explore a very common scenario when there is no labeled experimental data, and one must train a DL model using simulated or synthetic data. The trained model is then used to obtain predictions for the experimental data that typically contain structures and distortions not (fully) considered in the simulations. A similar approach can be applied to situations where the DL model(s) are trained on data from one experiment and applied to new previously unseen experimental images obtained under different conditions.

Our goal is to categorize every pixel in an image as belonging to an atom (or a particular type of atom) or to a background. In this case, the semantic segmentation of experimental atom-resolved images removes all the noise and returns a set of well-defined blobs (corresponding to atoms) on a uniform background where the centers of the segmented blobs correspond to atomic positions. The overall approach is summarized in Algorithm 1 and schematically depicted in Figure 1a. It starts with using simulated data to train ensemble of models instead of just a single model. Recent works have demonstrated that classical and Bayesian ensemble learning can lead to improved robustness of DL models and provide meaningful uncertainty quantification under the dataset shift and (potentially) for out-of-distribution data.[49, 50] Here, we combined classical ensemble learning,[51] where each new model is trained with different random initialization of weights and different random shuffling of training data, with a stochastic weight averaging (SWA) procedure,[52] which averages multiple points along the trajectory of stochastic gradient decent at the end of training. This approach is similar to multi-SWA and multi-SWA-Gaussian approaches described by Wilson *et al*.[50] but uses a constant learning rate and early stopping instead of a cyclic learning rate. Generally, we found that for simulation-to-real-world-transitions (i.e., model trained on simulated data is applied to experimental data), the role of ensemble learning is to identify an artifact-free model or a subset of models, that is, a model or models that do not "find" unphysical features in experimental data or at least a portion of experimental data. For experiment-to-experiment transitions (i.e., model trained on a subset of experimental data is applied to the remaining data), the ensemble learning can be used for improved robustness and for providing uncertainty estimates of predicted values for each point (pixel).

Each trained ensemble model is then applied to experimental data and the best (artifact-free) model(s) is selected by a human operator (a domain expert). For experimental data with high variability within the *xy* plane (single image) and/or along the time dimension (movie), there usually tend to be a portion of data for which the selected model(s) shows a sufficiently high detection rate (percentage of identified atoms). This portion of data (with additional pre-processing if necessary) is utilized to generate a new training set, this time from experimental data, that is used to train a new ensemble of models. This process can be repeated multiple times until the detection rate becomes sufficiently high for the entire dataset. Interestingly, we found that the addition of a small amount of artificial noise and distortions to the training data helps increase the detection rate of the retrained models on the remaining portion of experimental data, especially for dynamic data when there are significant structural changes between the first and the last movie frames.

**Algorithm 1** Steps to train in ELIT framework
1: $E_{initial} \leftarrow EnsTrain(D_{sim})$   ▷ Train ensemble of models
2: $P_{Ens} \leftarrow EnsPredict(E_{initial}, D_{exp})$   ▷ Apply ensemble to expt data
3: $E_0 \leftarrow ArtifactFree(E_{initial}, P_{Ens})$   ▷ Choose artifact-free models
4: **for** i in $\{1, \ldots, N\}$ **do**
5:   $D_i \leftarrow TrainData(E_{i-1})$   ▷ Generate training data
6:   $E_i \leftarrow EnsTrain(D_i)$   ▷ Train a new ensemble of models
7:   $P_{Ens} \leftarrow EnsPredict(E_{recent}, D_{exp})$   ▷ Apply ensemble to expt data
8: **done**
   **Optional Steps:**
9: $D_{MC} \leftarrow MultiClassData(E_N)$   ▷ Generate multi-class training set
10: $E_{final} \leftarrow EnsTrain(D_{MC})$   ▷ Train a new ensemble of models

**Algorithm 1.** Pseudocode for training models the ELIT framework. First (Line 1) an ensemble of models is trained on simulated data followed by applying each trained ensemble model (Line 2) to experimental data. The best (artifact-free) model or a subset of the best (artifact-free) models is then chosen (Line 3) and training data from a "good" portion of predictions from the best model(s) is generated. The new training data is used for training a new ensemble of models and make predictions on the experimental data using the newly trained ensemble (Lines 4-8), which in turn is used to refine the experimental training set for iterative re-training of the ensemble. As optional steps (Line 9, 10), a multi-class training set can be generated and used to train a new ensemble of models for multi-class classification.

Here, we have chosen two specific cases to showcase the efficacy of the ELIT framework for atom identification: (i) A dynamically evolving 2D graphene system and (ii) Lanthanum strontium manganite with embedded islands of nickel oxide (NiO-LSMO). The graphene movie contains multiple image frames, accounting for the dynamic nature of the structural evolution occurring in the graphene system with point impurities, whereas the NiO-LSMO image is a single image showing the presence of multiple type of atoms and phases.

We start by training an ensemble of 20 models using multi-SWA method for U-Net[40] model architecture. The training data represent 2D images with atoms as 2-dimensional Gaussians and corresponding circular masks of the fixed radius. Alternatively, the Multislice method[53] can be used for image simulations, although we did not find any significant differences within the context of the DL applications to 2D systems. The images and masks are generated using coordinates produced by ab initio molecular dynamics (AIMD) simulations performed on a graphene supercell at 300 K (more details in the Materials and Methods section). During each run of MD simulation, carbon atoms are removed one by one until a few (10) are left in the supercell. Once created, the images and masks are augmented via random cropping, application of Gaussian and Poisson noises with random scales, blurring, varying contrast levels, and zooming-in randomly in the range between 1x and 2x. The augmentation is used to help generalize to real experimental data. Examples of few snapshots of AIMD simulations and how images/masks are augmented are shown in the Supplementary Materials.

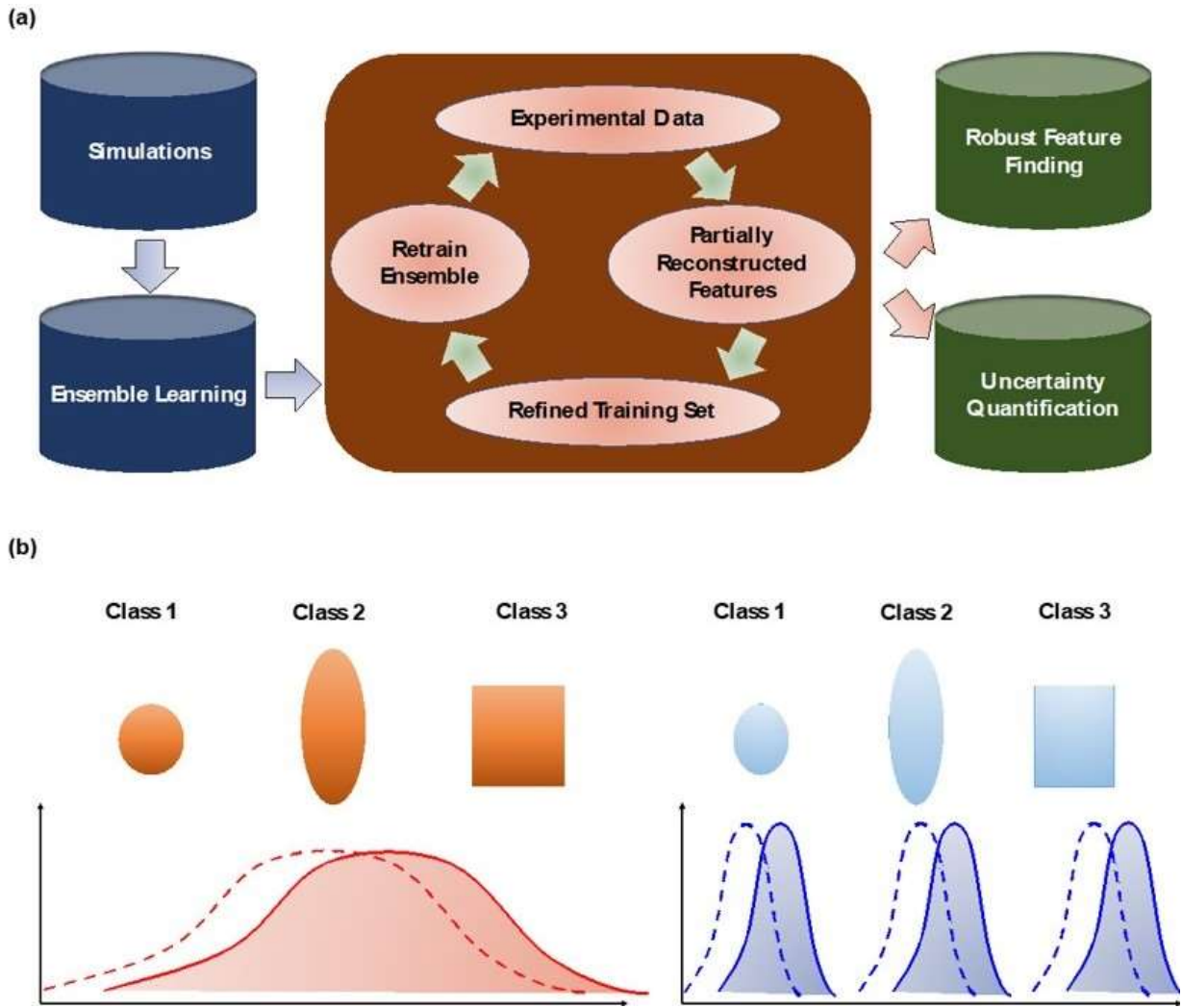

**Figure 1**. High-level schematic of the ELIT approach is shown in (a). (b) In a classically trained network, the training set has to be broad enough to allow sufficient variability to detect classes 1 and 2 presents in the system. However, a sufficiently broad training set will also potentially misidentify physically impossible class 3, leading to the classical balance between over- and underfitting. In the ELIT workflow, the multiple trained networks are used to identify the features present in the system, and retrain the network using these features. This allows "focusing" the attention of the network on the classes present in the system, thus dramatically improving the recognition.

The results of applying several selected models from the trained ensemble to the experimental data are shown in Figure 2 in a form of semantically segmented maps corresponding to the final layer of individual DL networks. We note that the accompanying interactive Jupyter notebook allows applying all the models in the ensemble to any experimental movie frame. Interestingly, we found a very significant variation in predictions on the experimental data despite that all the models showed nearly the same prediction accuracy (73.7±0.3%) on the simulated test data (see Supplementary Table I). This stark contrast in behavior on the theoretical test data and on real-world data is due to the fact that the former comes from the same distribution while the latter is almost always from a different distribution simply because it is impossible to account for all the instrumental parameters when preparing simulated data. One can see that the first (Fig. 2d-f) and second (Fig. 2g-i) models appear to be capable of avoiding amorphous regions even though the ensemble models were not specifically trained to do so (there was no amorphous phase in the training set). Notice, however, that the second model shows additional "foggy" features in the extended regions of graphene lattice, which do not have a direct physical interpretation and are considered an artifact. The third model does not automatically filter out the amorphous parts but tends to provide more clear results around point impurities. Finally, all the models in the ensemble start producing unphysical features inside the growing hole in graphene (Fig. 2e-f, 2h-i, 2k-l), which means that none of the current ensemble models can be used to analyze the entire dataset. Finally, we note that one cannot simply use an averaged ensemble prediction at this stage, as is common in classical DL, since it will average together both "physical" and "unphysical" features in predictions. By the same token, calculating the dispersion in the ensemble predictions, that is, obtaining uncertainty estimates for each point, is meaningless at this stage.

Next, we select the best, or one of the best, ensemble models (as defined by a domain expert) and create a new training set using the artifact-free subset of predictions on the experimental data. Alternatively, one can use an average prediction from a subset of the best ensemble models. For the graphene dataset, the subset of predictions used to create the new training data set is a portion of the movie (first 6 frames) without a hole. We use the third model from Figure 2 as our "baseline" model for the iterative training because it has a higher detection rate for point impurities. The features associated with the amorphous parts in the model output are removed via a patch-based Gaussian mixture model (GMM) analysis, where one first forms a stack of small image patches (here, we used a 48 x 48 patch size) around each identified "atom" acting as local descriptors and then applies a two-class GMM to the formed stack (see the Supplementary Materials and the accompanying notebook). Due to drastic differences in local neighborhoods of features associated with graphene and those associated with the amorphous part, the GMM unmixing easily separates a class corresponding to the graphene lattice and a class corresponding to the amorphous regions. The features associated with the latter are discarded during the new training set preparation. The new training data set is augmented by random cropping of experimental image frames and the constructed mask into 224 x 224 patches. In addition, we applied random $\pm 90^0$ rotations, horizontal and vertical flips, small scale jittering, and Gaussian noise with a randomly chosen magnitude. No contrast, blurring or other types of noise/distortions were applied to training data. The new ensemble of model can be trained using the best model from previous ensemble as a baseline, which can speed up the computation, or entirely from scratch. For the datasets studied in the current work, both strategies produced comparable results.

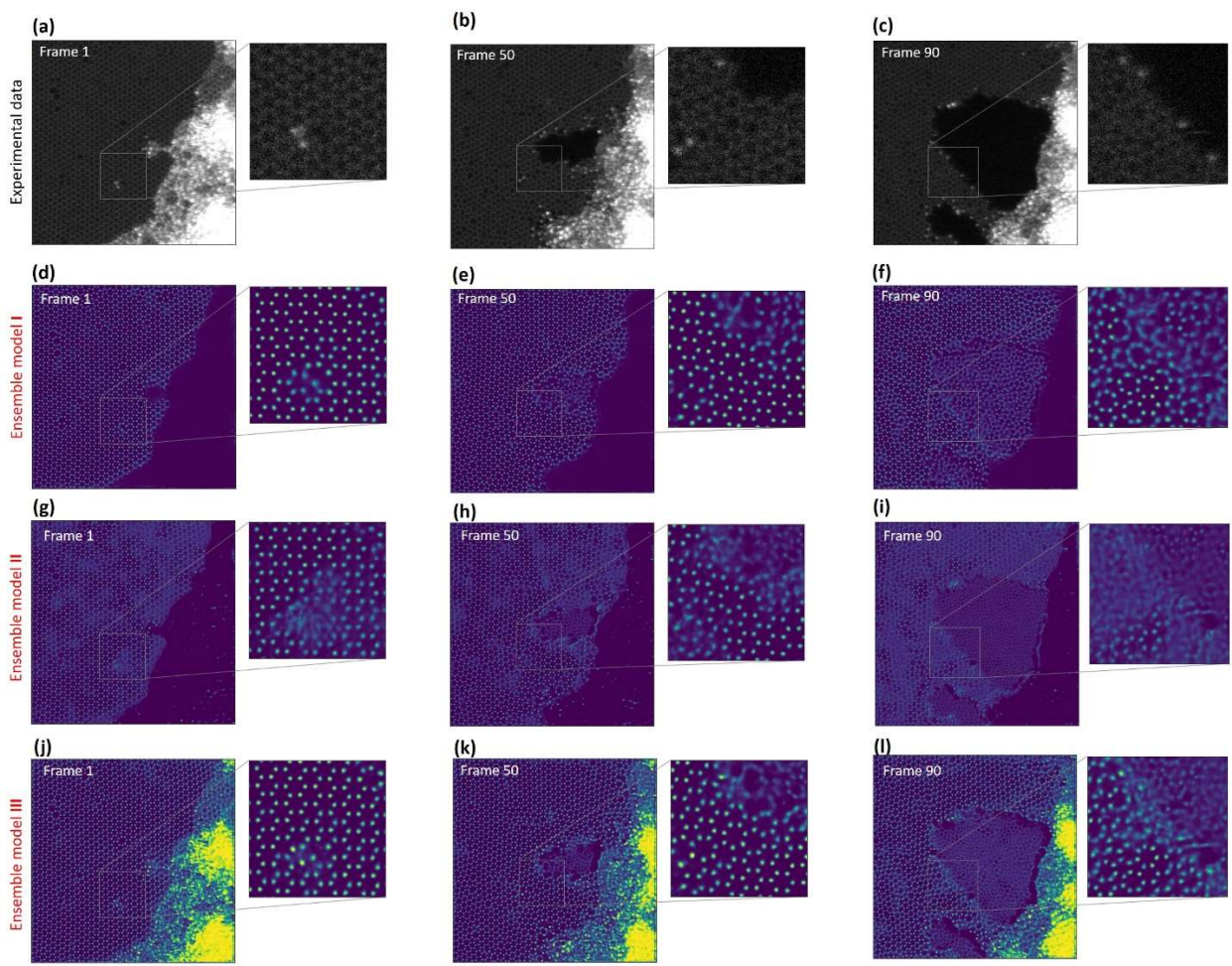

**Figure 2.** Application of ensemble models trained on simulated data to real data. Predictions in a form of semantically segmented maps from different ensemble models for frames 1 (a, d, g, j), 50 (b, e, h, k), and 90 (c, f, i, l) of the experimental movie from 2D graphene with impurities.

A new ensemble of model trained on the (augmented) subset of experimental data is then applied to the entire dataset. This time, since both training and test/validation data come from the same distribution, the variation in predictions of different ensemble models is much smaller and we can use a mean ensemble prediction instead of predictions of individual models, which as in the case of classical DL ensembles provides a more accurate prediction. The results for the frames from the middle and end of the movie are shown in Figure 3a, and 3b. Comparison of these results with the results of the models trained on simulated data (see Fig. 2) shows a significant improvement in the quality of predictions after just a single ELIT iteration. We can further add multiple channels to our training masks to allow for DL-based identification of both positions *and* type of atoms. Here, we used a simple binary threshold for separating intensities extracted at the atomic positions from the mean ensemble prediction. The resultant masks have 2 classes: graphene atoms and point impurities (here, the "impurity" is an atom with intensity higher than C). These masks and the corresponding experimental images are used to train another ensemble of models, now using randomly sampled patches from all the frames and applying the same augmentation procedures as in the previous iteration. The mean predictions of the ensemble of multi-class segmentation models are shown in Figure 3c, and 3d. The multi-class ensemble can be re-used on similar experimental data, that is, data obtained under similar experimental conditions with roughly the same feature scale (approximate number of pixels per atom).

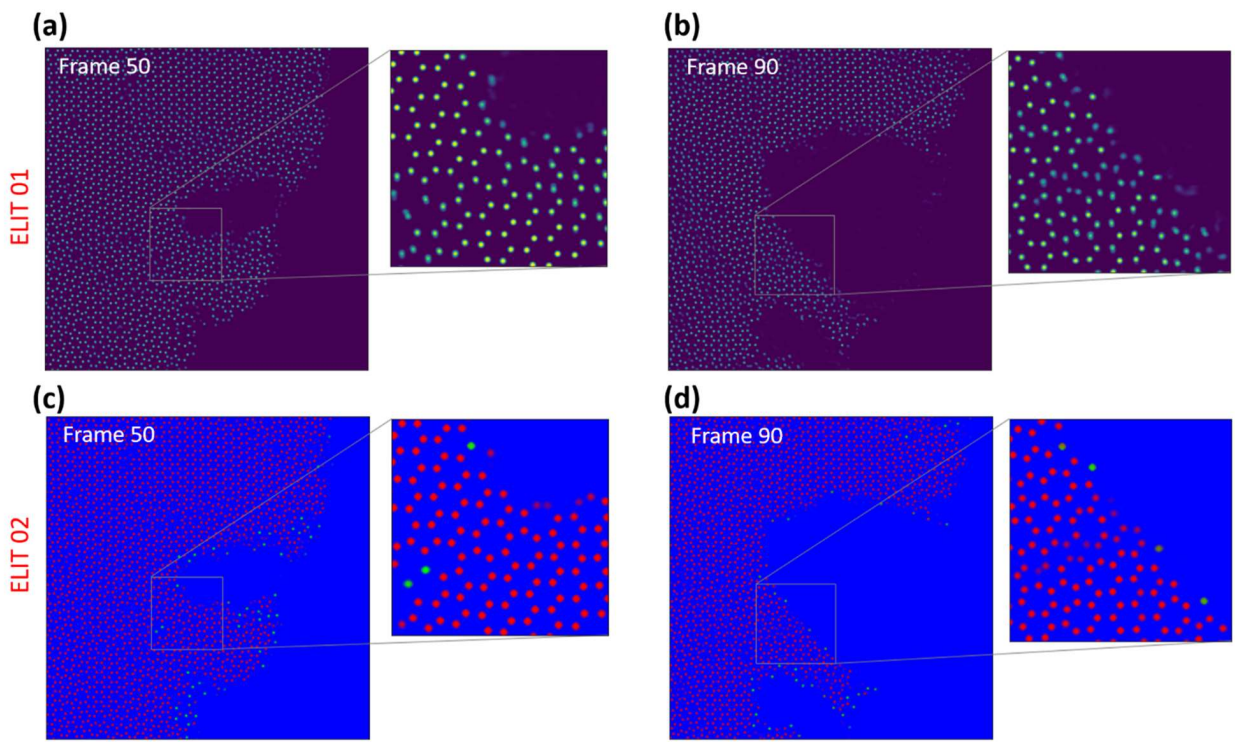

**Figure 3. Mean ensemble predictions after re-training with subsets of experimental data.** (a, b) Mean DL prediction (semantically segmented map) from the ensemble after the first re-training using only frames from the beginning of the movie. (c, d) Prediction (semantically segmented map) after second retraining using multi-class segmentation masks with training data randomly sampled from the entire movie.

Next, we demonstrate that ELIT framework can provide meaningful uncertainty estimates on the level of individual pixels. In Figure 4, we show the mean ensemble predictions for two sample regions and the corresponding uncertainty for each class computed as a standard deviation of ensemble predictions for each pixel. The larger uncertainty values at the boundaries of predicted blobs are expected as the (fixed) size of "atoms" in the mask is chosen somewhat arbitrary. At the same time the presence of "doubling" in uncertainty maps for impurities is interesting and can be related to unstable atoms jumping between different lattice sites during the scan. Note that this information cannot be captured by a single DL model, which would assign an atom to one of the positions without providing the uncertainty maps. We found that about 6-8 SWA models in the ensemble are typically enough for the meaningful pixel-level uncertainty estimation.

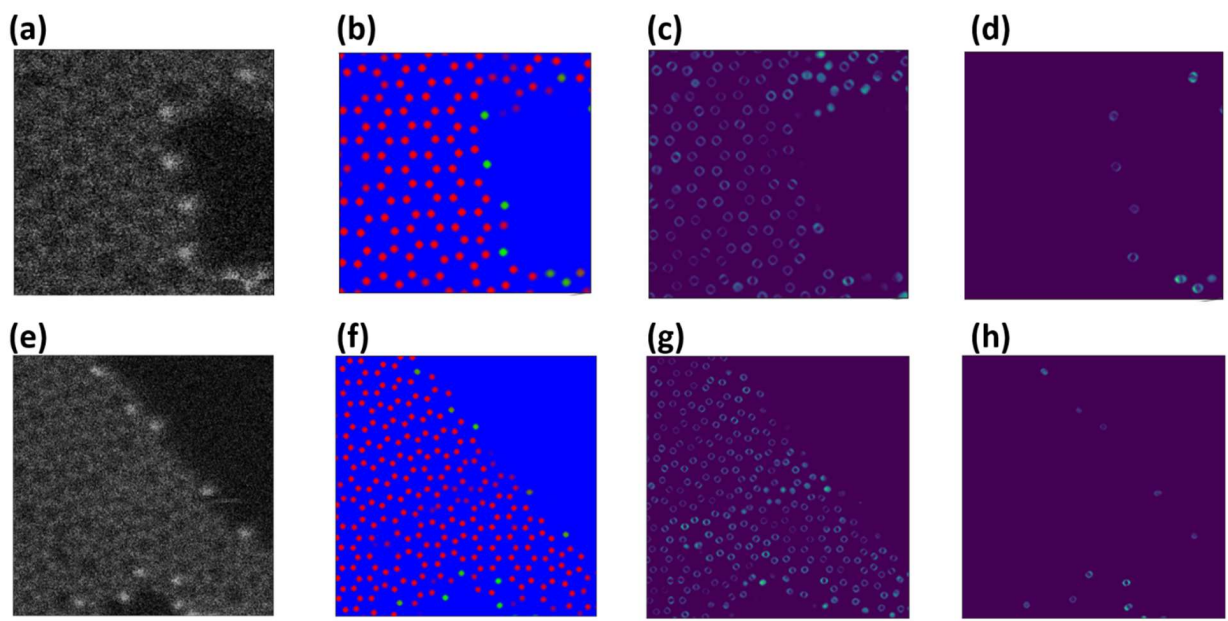

**Figure 4. Estimation of uncertainty in ensemble predictions on the level of individual pixels.** (a, e) Experimental sub-images from frames 95 and 90. (b, f) Mean ensemble prediction. (c, g) and (d, h) show normalized uncertainty maps calculated as a standard deviation of ensemble predictions in each pixel for graphene atoms (c, g) and impurity atoms (d, h). See the accompanying notebook for uncertainty maps on the full images. Higher intensity in color in (c.d) and (g,h) correspond to higher uncertainty associated with the atoms identified in those specific locations.

Finally, we show that the same approach can be used for finding and classifying (nearly) all the atoms in a single image using STEM data from a LSMO system with embedded NiO nano islands as an example (Fig. 5a). Here, a first ensemble of models is trained using a generic training set from a simple cubic lattice with random atomic displacements. When applied to real experimental data its models mostly fail to identify the second (NiO) structure and tend to "find" some unphysical features (Fig. 5b). Here we first used the best ensemble model to retrain an ensemble of models on the *same* simulated data using the weights of the best model as a baseline. This allowed for uncertainty estimation in the model's prediction that was used to select only the robust atomic features in the prediction, i.e., features associated with high variance in uncertainty maps were filtered out. The atomic positions associated with the remaining low-variance features were used to construct a new training set from the experimental data, which provided a significant improvement in the quality and rate of atomic segmentation/detection (Fig. 5c). Next, one may continue re-training a single class model until achieving a perfect detection rate (Fig. 5d) or switch to a multi-class classification as described earlier to categorize multiple sublattices in the system (Fig. 5e).

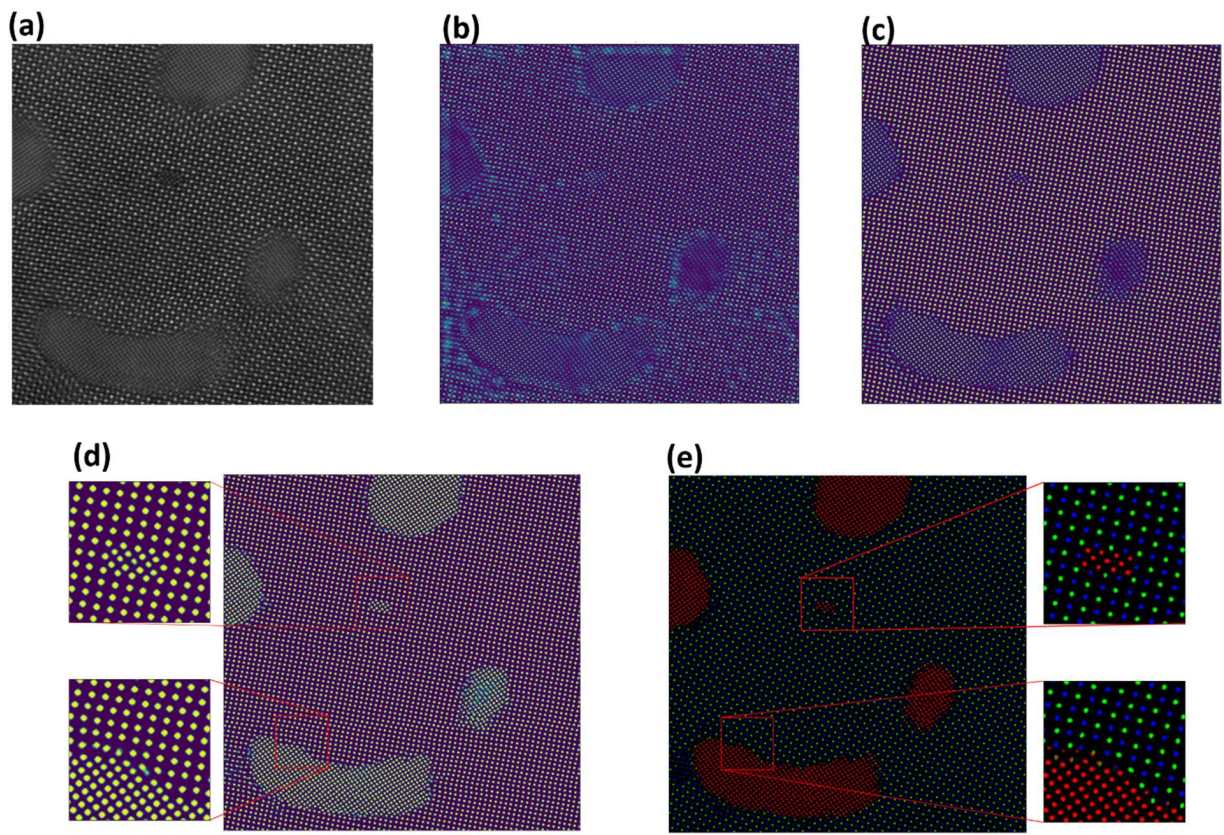

**Figure 5. Application of ELIT to a single image.** (a) Experimental STEM image of NiO-LSMO. (b) Result (semantically segmented map) from ensemble of models trained on simulated data. (c) Prediction (semantically segmented map) after the first ELIT re-training. (d, e) Predictions (semantically segmented maps) after the second ELIT retraining for a single class (d) and multiple classes (e).

In summary, we have developed the ELIT framework for iterative identification of features and corresponding pixel-wise uncertainty maps from high-resolution STEM images and demonstrated its application for several systems. We found that when applying model(s) trained on simulated data to real experimental data the ensemble learning can be used for selecting a subset of "physical" (artifact-free) models. These models can be iteratively retrained on "good" portions of experimental data in which case the new ensembles can be used for improving robustness of predictions and providing meaningful uncertainty estimations on the level of individual pixels. The method can be extended to other than U-Net deep learning architectures as well as to different imaging techniques.

This approach allows for two significant advances in DL applications to experiments. The first is the potential of ensemble-based uncertainty quantification. For the cases explored here and in the author's experience, uncertainties are often associated with unusual phenomena and can serve as a flag for unexpected behaviors. Perhaps even more importantly, this approach enables rapid correction for out of distribution drift during automated experiments where the imaging conditions change compared to the training set. Namely, it enables substituting the slow labeling/retraining of the network to rapid human-based deselection of the network from an ensemble.

**Acknowledgements**. This effort (machine learning) is based upon work supported by the U.S. Department of Energy (DOE), Office of Science, Office of Basic Energy Sciences Data, Artificial Intelligence and Machine Learning at DOE Scientific User Facilities (A.G., S.V.K.) and was also supported (STEM experiment) by the DOE, Office of Science, Basic Energy Sciences (BES), Materials Sciences and Engineering Division (O.D.), and was performed and partially supported (M.Z., BGS) at the Oak Ridge National Laboratory's Center for Nanophase Materials Sciences (CNMS), a DOE Office of Science User Facility. Dr. Matthew Chisholm (ORNL) is gratefully acknowledged for the STEM data on Ni-LSMO used in this work.

**Materials and methods**

*Samples preparation*

Atmospheric pressure chemical vapor deposition (AP-CVD) was used to grow graphene on Cu foil.[54] A coating of poly(methyl methacrylate) (PMMA) was spin coated over the surface to protect the graphene and form a mechanical stabilizer during handling. Ammonium persulfate dissolved in deionized (DI) water was used to etch away the Cu foil. The remaining PMMA/graphene stack was rinsed in DI water and positioned on a TEM grid and baked on a hot plate at 150 °C for 15 min to promote adhesion between the graphene and TEM grid. After cooling, acetone was used to remove the PMMA and isopropyl alcohol was used to remove the acetone residue. The sample was dried in air and baked in an Ar-$O_2$ atmosphere (10% $O_2$) at 500 °C for 1.5 h to remove residual contamination.[55] Before examination in the STEM, the sample was baked in vacuum at 160 °C for 8 h.

The LSMO-NiO VAN and the single-phase LSMO and NiO films were grown on STO (001) single-crystal substrates by PLD using a KrF excimer laser ($\lambda$= 248 nm) with fluence of 2 J/cm2 and a repetition rate of 5 Hz. All films were grown at 200 mTorr O2 and 700 °C. The films were post-annealed in 200 Torr of O2 at 700 °C to ensure full oxidation, and cooled down to room temperature at a cooling rate of 20 °C/min. For out-of-plane transport measurements, the films were grown on 0.5% Nb-doped STO (001) single-crystal substrates. The film composition was varied by using composite laser ablation targets with different composition.

*STEM imaging*

The plan-view STEM samples of Ni-LSMO were prepared using ion milling after mechanical thinning and precision polishing. In brief, a thin film sample was firstly ground, and then dimpled and polished to a thickness less than 20 micrometer from the substrate side. The sample was then transferred to an ion milling chamber for further substrate-side thinning. The ion beam energy and milling angle were adjusted towards lower values during the thinning process, which was stopped when an open hole appeared for STEM characterization. The STEM used for the characterization was a Nion UltraSTEM200 operated at 200 kV. The beam illumination half-angle was 30 mrad and the inner detector half-angle was 65 mrad. Electron energy-loss spectra were obtained with a collection half-angle of 48 mrad.

For graphene imaging, a Nion UltraSTEM 200 was used, operated at 100 kV accelerating voltage with a nominal beam current of 20 pA and nominal convergence angle of 30 mrad. Images were acquired using the high angle annular dark field detector.

*Ab Initio Molecular dynamics (AIMD)*

Details on AIMD simulations for the graphene movies: Ab-initio quantum-mechanical MD simulations were performed using the projector augmented plane-wave (PAW) method and PAW-PBE potential[56] as implemented in the Vienna ab initio simulation package (VASP).[57, 58]

A graphene supercell of 199 atoms with lattice parameters a = b = 24.68 Å, c = 8.60 Å with α = β = 90°, γ = 59.99° was considered for performing the simulations. All computations were carried out with a 400-eV plane-wave cutoff energy with appropriate Monkhorst Pack[59] k-point meshes, at 300 K temperature with 2000 timesteps of 1 fs each. Atoms were randomly removed from the supercell one-by-one and the MD simulation was repeated until 10 atoms remained in the supercell. The final converged coordinates for every timestep as produced by all the simulations were then used to prepare training set for building ensemble of models.

For NiO-LSMO, a generic square lattice with random displacements was utilized to simulate the trajectories, as used for training ensemble of models in the first step.

Data analysis:

All the deep learning routines were implemented using a home-built open-source software package AtomAI (https://github.com/pycroscopy/atomai) and are available via an interactive Jupyter notebook at https://github.com/aghosh92/ELIT

# Supplementary Materials

# Ensemble learning and iterative training (ELIT) for image analysis: applications to atom-resolved microscopy

Ayana Ghosh,[1,2] Bobby G. Sumpter,[1] Ondrej Dyck,[1]

Sergei V. Kalinin[1,2] and Maxim Ziatdinov,[1,2]

[1] The Center for Nanophase Materials Sciences, Oak Ridge National Laboratory, Oak Ridge, TN 37831

[2] Computational Sciences and Engineering Division, Oak Ridge National Laboratory, Oak Ridge, TN 37831

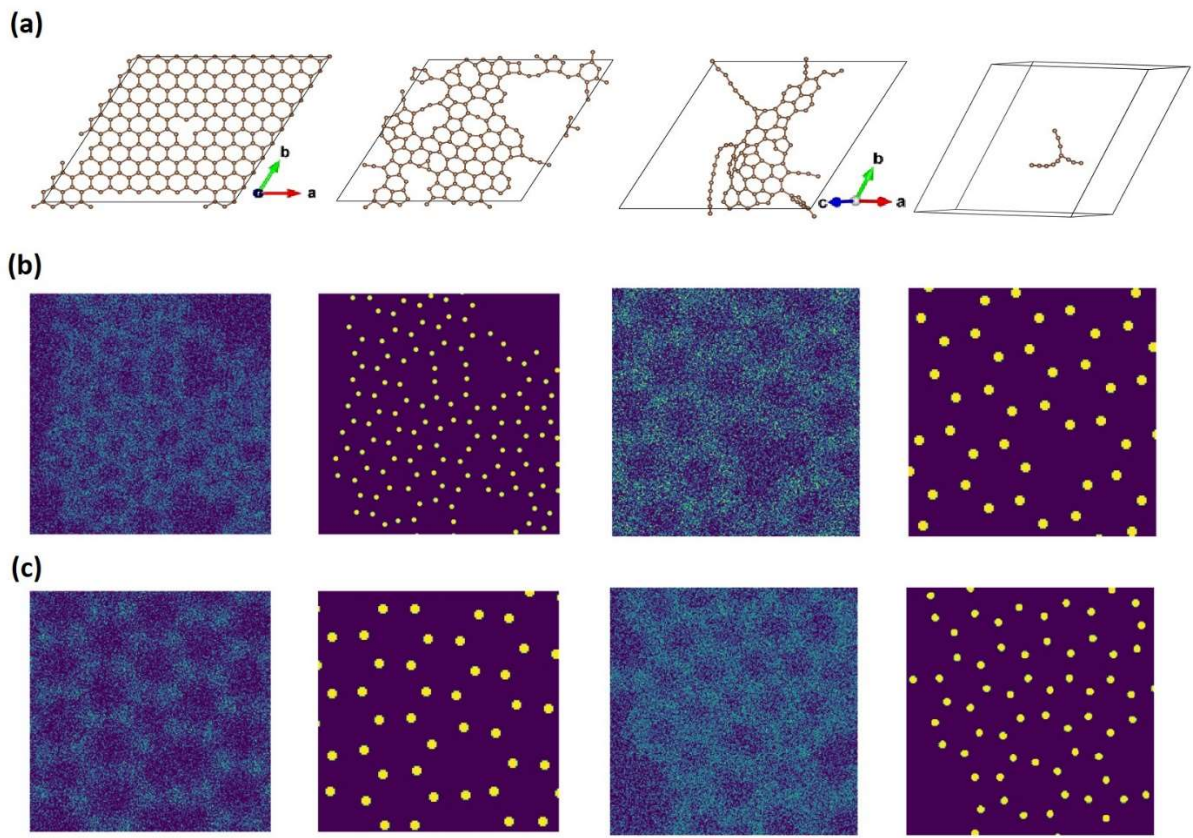

**Figure 1.** Snapshots from MD simulations are shown in (a) as performed on graphene supercell when 0, 50, 100 and 191 atoms are removed respectively. Several augmented image-mask pairs generated using the atomic coordinates from MD simulations and used for training the initial ensemble of models are displayed in (b-c).

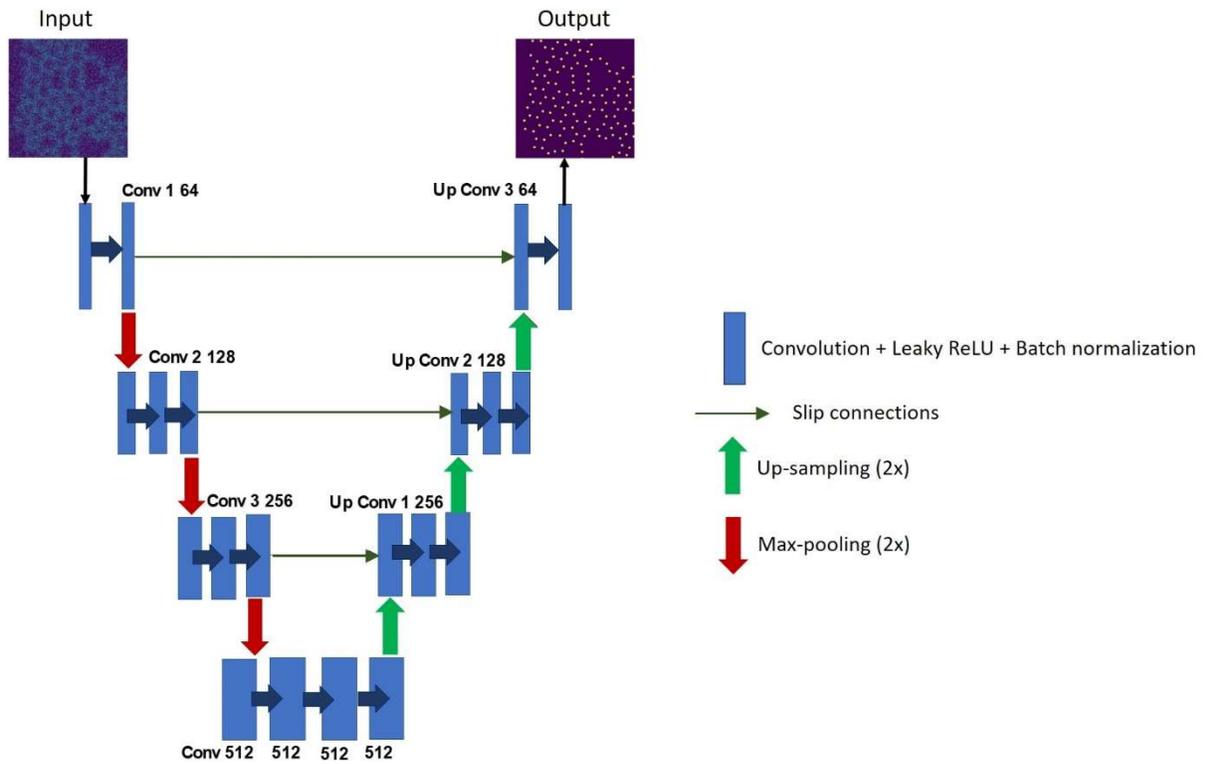

**Figure 2**. A schematic of the U-Net neural network architecture used in this work to construct ELIT models.

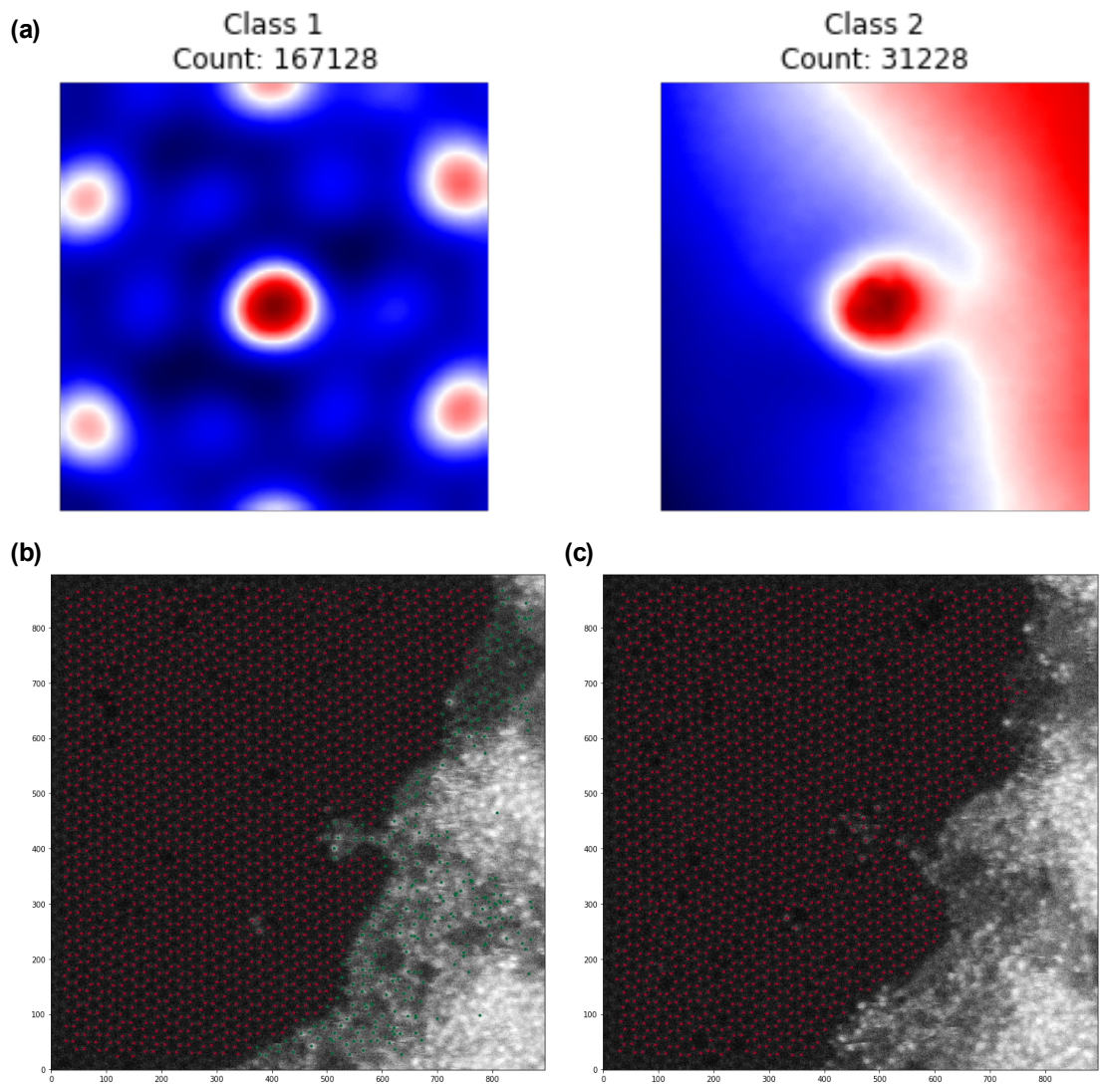

**Figure 3**. Example showing patch-based Gaussian mixture model (GMM) analysis, where one first forms a stack of small image patches (here, we used a 48 x 48 patch size) around each identified "atom" acting as local descriptors and then applies a two-class GMM to the formed stack. Here, the GMM unmixing easily separates a class corresponding to the graphene lattice (first class in (a) and red dots in (b), (c)) from the class corresponding to the amorphous regions (second class in (a) and green dots in (b)).

| Graphene Movie Predictions using pre-trained ensemble of models trained on MD data ||||
|---|---|---|---|
| Ensemble Model | IoU Score | Ensemble Model | IoU Score |
| 1 | 0.7327 | 11 | 0.7362 |
| 2 | 0.7336 | 12 | 0.7348 |
| 3 | 0.7382 | 13 | 0.7340 |
| 4 | 0.7361 | 14 | 0.7407 |
| 5 | 0.7404 | 15 | 0.7358 |
| 6 | 0.7320 | 16 | 0.7406 |
| 7 | 0.7402 | 17 | 0.7340 |
| 8 | 0.7334 | 18 | 0.7380 |
| 9 | 0.7384 | 19 | 0.7323 |
| 10 | 0.7397 | 20 | 0.7400 |
| Average | 0.7366 |||

**Table 1**. Intersection over Union (IoU) scores after individually applying 20 models in the ensemble trained on simulated data to a test set (also from simulated data).